\documentclass[article,11pt]{article}
\usepackage{jheppub}

\usepackage{caption}
\usepackage{subcaption}

\usepackage{amssymb}
\usepackage{amsbsy}
\usepackage{amsfonts}
\usepackage{amsmath}
\usepackage{epsfig}
\usepackage{cancel}
\usepackage{color}
\usepackage{graphicx}
\usepackage{epstopdf}
\usepackage{float}

\def\bea{\begin{eqnarray}}
\def\eea{\end{eqnarray}}
\def\be{\begin{equation}}
\def\ee{\end{equation}}
\def\beqn{\begin{eqnarray}}
\def\eeqn{\end{eqnarray}}
\def\beq{\begin{equation}}
\def\eeq{\end{equation}}

\def\Dslash{\not{\hbox{\kern-4pt $D$}}}
\def\pslash{\not{\hbox{\kern-4pt $p$}}}

\title{Electroweak precision measurements in supersymmetric models with a $U(1)_R$ lepton number}

\author[a]{Hugues Beauchesne}
\author[a]{Thomas~Gr\'egoire}

\emailAdd{gregoire@physics.carleton.ca}
\emailAdd{HuguesBeauchesne@cmail.carleton.ca}

\affiliation[a]{Ottawa-Carleton Institute for Physics, Department of Physics, Carleton University 1125 Colonel By Drive, Ottawa, K1S 5B6 Canada}
\abstract{ 
As experimental constraints on the parameter space of the MSSM and close variations thereof become stronger, the motivation to explore supersymmetric models that challenge some of the standard assumptions of the MSSM also become stronger. For example, models where the gauginos are Dirac instead of Majorana have recently received more attention. Beside allowing for a supersoft SUSY breaking mechanism where the gauginos only provide finite threshold corrections to scalar masses, the cross section for the production of a squark pairs is  reduced. In addition, Dirac gauginos can be used to build models that possess a $U(1)_R$ symmetry. This symmetry can then be identified with a lepton number, leading to models that are quite different from conventional scenarios. The sneutrinos in these models can acquire a {\it vev} and give mass to the leptons and the down-type squark. The phenomenology is novel, combining signatures that are typical of R-parity violating scenarios with signatures arising from leptoquarks. Correspondingly the constraints from electroweak precision data are also different. In these models, one of the leptons mixes with gauginos and superpotential Yukawa couplings can contribute to EWPM at tree level. In addition, lepton universality is broken. In this paper we adapt the operators analysis of Han and Skiba \cite{Han:2004az} to include the relevant violation of lepton universality, and do a global fit of the model to electroweak precision data, including all relevant tree-level and loop-level effects. We obtain bounds on the {\it vev} of the sneutrino and on the superpotential couplings of the model.

}

\begin{document}

\maketitle

\section{Introduction}
Supersymmetry (SUSY) as a solution to the hierarchy problem is now, after two years of LHC data, under severe tension. This has motivated the exploration of a wide range of supersymmetric models,  relaxing several assumptions built into the conventional Minimal Supersymmetric Standard Model (MSSM). One possibility put forward is to have models where the gauginos are Dirac instead of Majorana \cite{Hall:1990hq,Randall:1992cq, Nelson:2002ca, Fox:2002bu,Chacko:2004mi, Kribs:2007ac, Benakli:2008pg, Choi:2008pi,Kribs:2010md,Abel:2011dc,Davies:2011mp,Kumar:2011np,Bertuzzo:2012su,Itoyama:2012a,Itoyama:2012b,Itoyama:2012c,Chakraborty:2013}. While this requires an enlarged field content, it could also be less constrained by LHC data due to the fact that gluinos could be heavier with the same fine-tuning if SUSY breaking is supersoft \cite{Fox:2002bu} and that squark pair production has a smaller cross-section when the gluinos are Dirac \cite{Heikinheimo:2011fk,Kribs:2012gx}. Also, with Dirac gauginos, one can impose an approximate $U(1)_R$ symmetry on the model  and this was shown to alleviate flavor constraints on the squark mass matrices \cite{Kribs:2007ac}. Furthermore, the $U(1)_R$ symmetry can be identified with a lepton number \cite{Gherghetta:2003he,Frugiuele:2011mh,Frugiuele:2012pe} (see \cite{Fayet:1974pd,Fayet:1976et,Fayet:1977yc,Fayet:1978qc} for earlier work) . This allows for reduction of the Higgs sector of the $U(1)_R$ model (which would otherwise need four doublets \cite{Kribs:2007ac}), by allowing the sneutrino to get a vacuum expectation value ({\it vev}) and play the role of the down-type Higgs. In this setup, the constraints on the {\it vev} of the sneutrino are much milder than in traditional supersymmetric models because the sneutrino does not carry lepton number and giving it a {\it vev} does not introduce unacceptably large neutrino masses. The down-type quarks and the leptons then acquire a mass through $R$-parity violating coupling (where here $R$-parity means the conventional $R$-parity, not the $Z_2$ subgroup of the full $U(1)_R$ we are imposing on the model). It is well known that such couplings will lead to deviations to electroweak precision observables (see \cite{Barbier:2004ez} for a review), and it is the goal of this paper to study more in depth the constraints that electroweak precision observables put on the model.  A preliminary study of such constraints was performed in \cite{Frugiuele:2011mh,Frugiuele:2012pe}, and LHC phenomenology was studied in \cite{Frugiuele:2012kp}. In \cite{Frugiuele:2011mh,Frugiuele:2012pe}, the constraints were estimated by looking at the tree-level effect of the new couplings which were considered in isolation. In this paper, we analyze the constraints more thoroughly by considering loop effects as well and fitting to all relevant electroweak data.  We adopt the operator method of \cite{Han:2004az} to analyze the constraints on the parameter space of the model. However, because our model leads to deviations from flavor universality in the lepton sector, we need to generalize the analysis of \cite{Han:2004az,Han:2005pr} to take these effects into account. 

\section{The Model}
In this section, we present a review of the model which was described in details in \cite{Frugiuele:2011mh,Frugiuele:2012pe}. In addition to the usual matter chiral superfields, $Q,U^c,D^c,L,E^c$, the model has two $SU(2)$ doublet chiral superfields:
\begin{equation}
H_u, R_d
\end{equation}
with hypercharge $+1/2$ and $-1/2$ respectively, and a set of chiral superfields that are in the adjoint representation of the Standard Model gauge group:
\begin{equation}
G, T, S
\end{equation}
where $G$ is an adjoint of color, $T$ an adjoint of $SU(2)$ and $S$ a singlet. 

The superfield $R_d$ has the same quantum numbers as the down-type Higgs superfield of the MSSM, but we do not use the name $H_d$ as in our model this field does not acquire a {\it vev}. Its scalar component is an inert doublet. The adjoint chiral superfields are added to give Dirac masses to gauginos.  These masses occur through a superpotential term of the form \cite{ Fox:2002bu} (shown here for the winos):
\begin{equation}
\int d^2 \theta \frac{W'^\alpha W^a_\alpha T^a}{M}
\end{equation}
where $W'$ is a real superfield that acts as a spurion with a non-zero $D$ term: $W'_\alpha = D' \theta_\alpha + \cdots$.  In addition to a mass term of the form
\begin{equation}
\label{eq:diracmass}
M_2^D \lambda^a \Psi^a ,
\end{equation}
the superpotential term above also leads, once the classical equations of motion are solved, to a non-standard expression for $D$:
\begin{align}\label{Dequation}
D^a  =& g \left(H_u^\dagger \tau^a H_u + R_d^\dagger \tau^a R_d + L^\dagger \tau^a L + Q^\dagger \tau^a Q + T^\dagger \lambda^a T \right) + \sqrt{2}(M_2^D T^a + \text{h.c.})\\ \nonumber
D_Y =& g' \sum_\Phi Y_{\Phi} \Phi^\dagger \Phi + \sqrt{2} (M_1^D S + \text{h.c.})
\end{align}
where $\tau^a$ and $\lambda^a$ are respectively the generators of the two and three dimensional representations of $SU(2)$. 

As mentioned above, it is the sneutrino and not $R_d$ which acquires a {\it vev} providing masses to down-type fermions in this model. Therefore, the part of the superpotential responsible for the masses of the standard model particles will be of the form:
\begin{equation}\label{W0}
W_0 = y_u Q H_u U^c + y_d Q L_a D^c + y_{e_b} L_a L_b E^c_b+y_{e_c} L_a L_c E^c_c + \mu H_u R_d .
\end{equation}
The subscript $a$ on the lepton chiral superfield denotes the flavor of the sneutrino that acquires a {\it vev}. The two other lepton flavors are labeled $b$ and $c$. This superpotential is invariant under  a $U(1)_R$ symmetry where the superfields $L_a$ and $H_u$ have R-charge 0, while $Q, U^c, D^c, L_b, L_c$ have $R$-charge 1 and $E^c_b, E^c_c$ and $R^d$ have $R$-charge 2.  The $Q L D^c$ and $ L L E^c$ terms are the standard R-parity violating terms, which in the literature have coefficients called $\lambda'$ and $\lambda$ respectively. Here they are the Yukawa couplings. We also include in the superpotential terms of the form:

\begin{equation}\label{Wadj}
W_{adj} = \lambda_S S H_u R_d + \lambda_T H_u T R_d,
\end{equation}
where $T=T^a\sigma^a$, which respect all the symmetries of the model and can contribute to raising the Higgs mass at loop level \cite{Bertuzzo:2014}.  
We take the soft SUSY breaking terms, in addition to the Dirac gaugino masses (see eq. \eqref{eq:diracmass}), to be:
\begin{equation}
V_{soft} = \sum_i m_i^2 \Phi_i^\dagger \Phi_i + \left[ \frac{1}{2} b_T T^2 + \frac{1}{2} b_S S^2 + B_\mu H_u L_a + \text{h.c.}\right],
\end{equation}  
where the sum runs over all scalars. We see here that because $R_d$ doesn't have a $B_\mu$-term, it will not get a {\it vev}, as long as its mass squared is positive. On the other hand, due to the presence of the $B_\mu$ term, the sneutrino ${\tilde \nu_a}$will acquire one. 

\section{Contributions to electroweak precision measurements}
In this section, the corrections to the electroweak precision measurements (EWPM) coming from the new physics are presented. Since all new particles are experimentally constrained to be rather heavy, their contributions can be parametrized to a good approximation in terms of dimension six effective operators respecting the Standard Model gauge symmetries. We use the same basis as in \cite{Han:2004az} but do not assume a full $U(3)^5$ flavor symmetry, which is not present in our model since only one flavour of sneutrino gets a {\it vev}. However, we do not consider operators that lead to FCNC in the lepton or quark sector as those would be much more strongly constrained, and can be avoided in our model by appropriate assumptions on the flavour sector. The operators containing only gauge bosons and Higgs particles are \cite{Han:2004az}
\begin{equation}\label{operators1}
		O_{WB}=(h^{\dagger}\sigma^{a}h)W^{a}_{\mu\nu}B^{\mu\nu}\qquad\qquad O_h=|h^{\dagger}D_{\mu}h|^2.
\end{equation}
The four-fermion operators that are relevant to EWPM, where we explicitly show the lepton flavor indices, are \cite{Han:2004az}
\begin{equation}\label{operators2}
	\begin{aligned}
	O^s_{ll}[mn] & =\frac{1}{2}(\overline{l}^{m}\gamma^{\mu}l^{m})(\overline{l}^{n}\gamma_{\mu}l^{n}) &
	O^t_{ll}[mn] & =\frac{1}{2}(\overline{l}^{m}\sigma^a\gamma^{\mu}l^{m})(\overline{l}^{n}\sigma^a\gamma_{\mu}l^{n})\\
	O^s_{lq}[m] & =(\overline{l}^{m}\gamma^{\mu}l^{m})(\overline{q}\gamma_{\mu}q) &
	O^t_{lq}[m] & =(\overline{l}^{m}\sigma^a\gamma^{\mu}l^{m})(\overline{q}\sigma^a\gamma_{\mu}q)\\
	O_{le}[mn] & =(\overline{l}^{m}\gamma^{\mu}l^{m})(\overline{e}^{n}\gamma_{\mu}e^{n}) &
	O_{qe}[m] & =(\overline{q}\gamma^{\mu}q)(\overline{e}^{m}\gamma_{\mu}e^{m})\\
	O_{lu}[m] & =(\overline{l}^{m}\gamma^{\mu}l^{m})(\overline{u}\gamma_{\mu}u) &
	O_{ld}[m] & =(\overline{l}^{m}\gamma^{\mu}l^{m})(\overline{d}\gamma_{\mu}d)\\
	O_{ee}[mn] & =\frac{1}{2}(\overline{e}^{m}\gamma^{\mu}e^{m})(\overline{e}^{n}\gamma_{\mu}e^{n}) &
	O_{eu}[m] & =(\overline{e}^{m}\gamma^{\mu}e^{m})(\overline{u}\gamma_{\mu}u)\\
	O_{ed}[m] & =(\overline{e}^{m}\gamma^{\mu}e^{m})(\overline{d}\gamma_{\mu}d).  \\
	\end{aligned}
\end{equation}
The operators that modify the coupling between fermions and gauge bosons are \cite{Han:2004az} \footnote{Because this is a two Higgs doublets model, one could also write operators with the second doublet. However, the effects of those on precision observables can be absorbed in the operators of (\ref{operators3}). }: 
\begin{equation}\label{operators3}
	\begin{aligned}
	O^s_{hl}[m] & =i(h^\dagger D^{\mu}h)(\overline{l}^{m}\gamma_{\mu}l^{m})+\mbox{h.c.}&
	O^t_{hl}[m] & =i(h^\dagger \sigma^aD^{\mu}h)(\overline{l}^{m}\sigma^a\gamma_{\mu}l^{m})+\mbox{h.c.}\\
	O^s_{hq} & =i(h^\dagger D^{\mu}h)(\overline{q}\gamma_{\mu}q)+\mbox{h.c.}&
	O^t_{hq} & =i(h^\dagger \sigma^aD^{\mu}h)(\overline{q}\sigma^a\gamma_{\mu}q)+\mbox{h.c.}\\
	O_{he}[m] & =i(h^\dagger D^{\mu}h)(\overline{e}^{m}\gamma_{\mu}e^{m})+\mbox{h.c.}&
	O_{hu} & =i(h^\dagger D^{\mu}h)(\overline{u}\gamma_{\mu}u)+\mbox{h.c.}\\
	O_{hd} & =i(h^\dagger D^{\mu}h)(\overline{d}\gamma_{\mu}d)+\mbox{h.c.} \\
	\end{aligned}
\end{equation}
 Finally, there is one  operator affecting only gauge boson self-interactions \cite{Han:2004az}
\begin{equation}\label{operators4}
	O_{\hat{W}}=\epsilon^{abc}W^{a\nu}_{\mu}W^{b\lambda}_{\nu}W^{c\mu}_{\lambda}.
\end{equation}
The total effective Lagrangian is therefore the sum of the Lagrangian of the SM and a linear combination of the different dimension six operators:
\begin{equation}
\mathcal{L} = \mathcal{L}_{\text{SM}} + a_i O_i
\end{equation}
where $O_i$ represent the operators, and $a_i$ are coefficients with dimension of inverse mass squared.

In the following sections, we present the contributions to the different operators. We first compute the coefficients of operators related to oblique corrections. Then, the four-fermion operators coming from scalar exchange and box diagrams are presented. Operators that modify gauge boson vertices coming from loop diagrams and mixing are shown. Finally, the contributions to the operator $O_{\hat{W}}$ are discussed. 

\subsection{Oblique parameters}
\begin{figure}[tbp]
		\begin{center}
			\includegraphics[scale=0.9, draft=false, trim=0.0cm 0.0cm 0.0cm 0cm, clip=true]{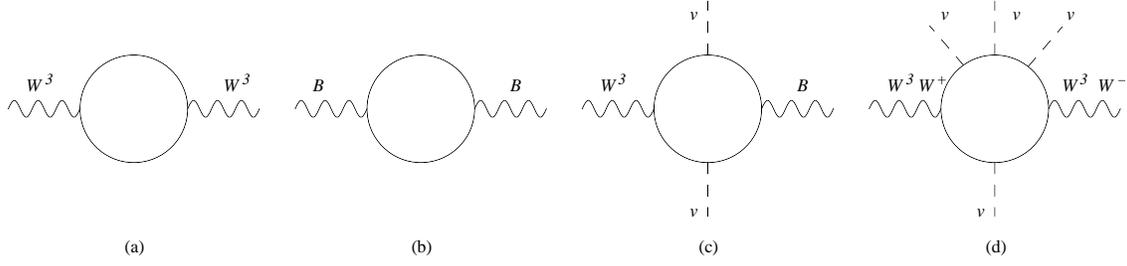}
		\end{center}
		\caption{\label{Obliquediagrams} Diagrams contributing to the oblique parameters. Plain lines correspond to unspecified superpartners or leptons $a$. (a), (b), (c) and (d) correspond to $W$, $Y$, $\hat{S}$ and $\hat{T}$ respectively. $v$ stands for the insertion of a {\it vev}. Reproduced from \cite{Marandella:2005wc}.}
\end{figure}

A standard and convenient way of parametrizing deviations to EWPM is through the so-called oblique parameters \cite{Peskin:1991sw,Peskin:1990zt}, which are defined to be modifications to two point functions of the electroweak gauge bosons. Coefficients of some of the higher dimensional operators mentioned above can in turn be written as a function of the oblique parameters. In this section, we present these relationships and the computation of the oblique parameters in our model due to a {\it vev} of the third component of the triplet $T_3$ and loops of scalars and fermions.  We use the definitions of \cite{Marandella:2005wc} for the definitions of the oblique parameters $\hat{S}$, $\hat{T}$, $Y$ and $W$ (see also \cite{Barbieri:2004qk}):
\begin{equation} 
	\hat{S}=\frac{g}{g'}\Pi'_{W_3Y}(0)\;\;\;\;\hat{T}=\frac{\Pi_{W_3W_3}(0)-\Pi_{W^{+}W^{-}}(0)}{M_W^2} 
	\;\;\;\; Y=\frac{M_W^2}{2}\Pi''_{YY}(0)\;\;\;\; W=\frac{M_W^2}{2}\Pi''_{W_3W_3}(0).
\end{equation}
Following  \cite{Marandella:2005wc}, we compute loop-level contributions to these parameters by considering diagrams of the form shown in figure $[$\ref{Obliquediagrams}$]$, where the {\it vev}s are treated as perturbations. The first two parameters are linked to $O_h$ and $O_{WB}$ by
\begin{equation}
	a_{WB}=\frac{g'\hat{S}}{gv^2}\qquad a_h=-\frac{2\hat{T}}{v^2}.
\end{equation}
The two others are related to operators which are not listed in \eqref{operators1}:
\begin{equation}
\label{eq:WYop}
	O_Y=\frac{(\partial_{\rho}Y_{\mu\nu})^2}{2} \qquad \qquad O_{W}=\frac{(D_{\rho}W_{\mu\nu})^2}{2}.
\end{equation}
However, using the equations of motions, these operators can be written in terms of the operators shown in \eqref{operators1}, \eqref{operators2} and \eqref{operators3} (plus additional operators that do not contribute to EWPM). Only eliminating $O_Y$ leads to additional contribution to the operators of \eqref{operators1}. More precisely it gives rise to an operator, amongst others, of the form
\begin{equation}
	\left(h^{\dagger}D^{\mu}h+\mbox{h.c.}\right)^2.
\end{equation}
This operator can then be related to $O_h$ and terms irrelevant to EWPM. The net effect on $a_h$ due to $O_Y$  is:
\begin{equation}
	a_{h}|_Y=-\frac{g'^2Y}{4M_W^2}.
\end{equation}
In practice, this contribution is overshadowed by the $\hat{T}$ term. The operator $O_W$  can also be eliminated using the equations of motion, but does not give a contribution to the operators of \eqref{operators1}.

\subsubsection{{\it Vev} of $T^3$}
In general, the field $T^3$ will acquires a {\it vev} of the form
\begin{equation}
\label{vevT3}
		v_{T^3}= \frac{ g M_2^D(v_u^2-v_a^2)- \sqrt{2}\lambda_T\mu v_u^2}{2 \left(m_{T_R}^2+\frac{(\lambda_T)^2v_u^2}{2}\right)},		
\end{equation}
where $m_{T_R}^2\equiv m_{T}^2+b_T+4 (M_2^D)^2$ and $v_u$ and $v_a$ are the {\it vev} of the neutral component of $h_u$ and $l_a$ respectively. This will give a tree-level contribution to $\hat{T}$ through a correction to the charged $W$ boson mass:
\begin{equation}
	\hat{T}_{v_{T^3}}=\frac{4v_{T^3}^2}{v^2}.
\end{equation}
For a light enough scalar triplet, this contributions dominates the $\hat{T}$ parameter. The singlet also acquires a small {\it vev}. However, it is not a direct contribution and its effect is therefore much smaller and can be ignored.

\subsubsection{Scalars}
Loops of scalars will contribute to the oblique parameters. The relevant couplings come from \eqref{Dequation}, \eqref{W0} and \eqref{Wadj}. The first term of each line of \eqref{Dequation} are present in the MSSM and lead to contributions similar to those presented in \cite{Marandella:2005wc}. 

For sfermions, the results of \cite{Marandella:2005wc} apply almost directly. The only difference is that, as $l_a$ provides mass to the down type quarks and leptons, we include its scalar contributions in the Higgs sector. This has the advantage of making expressions simpler and easier to compare. The sfermions contributions are
\begin{equation}\label{Sfermions}
	\begin{aligned}
		W_{sfermions}&=\frac{\alpha_2 M_W^2}{80\pi}\left(\frac{2}{3}\frac{1}{m_L^2} +\frac{3}{m_Q^2}\right)\\
		Y_{sfermions}&=\frac{\alpha_Y M_W^2}{40\pi}\left(\frac{4}{3}\frac{Y_L^2}{m_L^2}+\frac{Y_E^2}{m_E^2}+3\frac{Y_D^2}{m_D^2} +3\frac{Y_U^2}{m_U^2}+6\frac{Y_Q^2}{m_Q^2}\right)\\
		\hat{S}_{sfermions}&=-\frac{\alpha_2}{24\pi}\left[M_W^2\left(2\frac{Y_L}{m_L^2}+9\frac{Y_Q}{m_Q^2}\right) +\frac{1}{2}\frac{m_t^2}{m_Q^2}\right]\\
		\hat{T}_{sfermions}&=\frac{\alpha_2M_W^2}{16\pi}\cos^2 2\beta\left(\frac{2}{3}\frac{1}{m_L^2}+\frac{2}{m_Q^2}\right)+\hat{T}_{stop},
	\end{aligned}
\end{equation}
where $\tan \beta = v_u\slash v_a$ and $\hat{T}_{stop}$ is given in \cite{Marandella:2005wc}:
\begin{equation}
	\hat{T}_{stop}=\frac{\alpha_2}{16\pi}\frac{\left(M_W^2\cos^2 2\beta+m_t^2\right)^2}{m_Q^2 M_W^2}.
\end{equation}	
The $R_d$ scalar field, despite not acquiring a {\it vev}, still contributes to the oblique parameters. For $Y$ and $W$, its contributions are similar to those of $l_b$ and $l_c$ and can be read from \eqref{Sfermions} directly. For $\hat{S}$ and $\hat{T}$, the presence of \eqref{Wadj} changes the result. The parameters are
\begin{equation}
	\begin{aligned}
		W_{R_d}&=\frac{\alpha_2 M_W^2}{240\pi M_{R_d}^2}\\
		Y_{R_d}&=\frac{\alpha_Y M_W^2}{240\pi M_{R_d}^2}\\
		\hat{S}_{R_d}&=\frac{\alpha_2M_W^2}{48\pi M^2_{R_d}}\left(\cos{2\beta}-\frac{2\sin^2\beta(\lambda_T^2-\lambda_S^2)}{g^2} \right)\\
		\hat{T}_{R_d}&=\frac{\alpha_2M_W^2}{48\pi M^2_{R_d}}\left(\cos{2\beta}-\frac{2\sin^2\beta(\lambda_T^2-\lambda_S^2)}{g^2} \right)^2,
	\end{aligned}
\end{equation}
where $M_{R_d}^2\equiv\mu^2+m_{R_d}^2$.
The Higgs sector (including the scalar part of $l_a$) gives contributions  to $\hat{S}$ ,$W$ and $Y$ that can once again be obtained by using the results of \cite{Marandella:2005wc} almost directly. In terms of $m_{A^{0}}^2=m_{H_u}^2+m_{L}^2+\mu^2$, these contributions are
\begin{equation}
	\begin{aligned}
		W_{Higgs}&=\frac{\alpha_2}{240\pi}\frac{M_W^2}{m_{A^{0}}^2}\\
		Y_{Higgs}&=\frac{\alpha_Y}{240\pi}\frac{M_W^2}{m_{A^{0}}^2}\\
		\hat{S}_{Higgs}&=-\frac{\alpha_2}{48\pi}\frac{M_W^2}{m_{A^{0}}^2}\left(1-\frac{M_Z^2}{2M_W^2}\sin^2 2\beta\right).
	\end{aligned}
\end{equation}
The presence of the scalar components of the singlet and triplet do not lead to any contribution to $Y$ and $\hat{S}$ by themselves, but only a contribution to $W$:
\begin{equation}
		W_{ scalar\,gauge}=\frac{\alpha_2 M_W^2}{120\pi}\left(\frac{1}{m_{T_R}^2}+\frac{1}{m_{T_I}^2}\right),
\end{equation}
where $m_{T_I}^2\equiv m_{T}^2-b_T$. 
The contribution to $\hat{T}$ from diagrams with higgs, triplet and singlets is however more difficult to compute using insertions of the Higgs {\it vev} because of the mixing between the Higgs and the triplet. Therefore, we compute these contributions by numerically diagonalizing the scalar mass matrix.

\subsubsection{Higgsinos and gauginos}	
The fit constrains the value of $v_a$ to a region of phase space where it is much smaller than $v_u$. As such, the contributions to the oblique parameters containing only powers of $v_u$ dominate and are presented here. The contributions containing powers of $v_a$ were also included in the numerical fit. With the exception of the contributions coming from couplings $\lambda_S$ and $\lambda_T$, the diagrams with binos are usually smaller by an order of magnitude or so and are not presented for simplicity, but were included in the fit. The dominant terms in the limit of $\lambda_S$ and $\lambda_T$ small are
\begin{equation}
	\begin{aligned}
		W_{fermions}&=\frac{\alpha_2}{30\pi}\left(\frac{4M_W^2}{(M_{2}^D)^2}+\frac{M_W^2}{\mu^2}\right)\\
		Y_{fermions}&=\frac{\alpha_Y M_W^2}{30\pi \mu^2}\\
		S_{fermions}&=\frac{\alpha_2 M_W^2}{12\pi (M_{2}^D)^2}\left[\frac{a(a-5-2a^2)}{(a-1)^4}+\frac{(1-2a+9a^2-4a^3+2a^4)}{(a-1)^5}\ln a\right]\sin^2\beta\\
		T_{fermions}&=\frac{\alpha_2 M_W^2}{48\pi (M_{2}^D)^2}\left[\frac{19-64a-91a^2+16a^3}{(a-1)^4}+\frac{6a(-4+25a-a^2)}{(a-1)^5}\ln a\right]\sin^4\beta,
	\end{aligned}
\end{equation}
where $a\equiv (\mu/M_{2}^D)^2$. The oblique parameters containing $\lambda_S$ and $\lambda_T$ are included but not presented here as the expressions are rather long. The contribution to the parameter $\hat{T}$ is in fact the dominant term in setting limits on $\lambda_S$ and $\lambda_T$ for massive enough scalar gauge particles.

\subsection{Four-fermion operators}
There are many contributions to the four-fermion operators. These can come from the operators \eqref{eq:WYop} associated to $W$ and $Y$ once the equation of motion are used, from  scalar exchange and from box diagrams. The contributions proportional to $W$ and $Y$ are given by:
\begin{equation}
	a^s_{ij}=-Y_iY_jg'^2\frac{Y}{2M_W^2} \qquad a^t_{ij}=-g^2\frac{W}{8M_W^2},
\end{equation}
where $i$ and $j$ stand for the different combinations of fields possible. The family indices are suppressed because they are the same for every combination, as is expected from the fact that the oblique parameters are universal.

The tree-level scalar exchange contributions come from the exchange of sfermions between leptons or down quarks. The scalars are integrated out and the diagrams are rewritten using the Fiertz rearrangement formulas. In practice, most of them have a negligible effect on the fit as they involve small Yukawa couplings. The only important ones are those which contribute to the observables $R_{\tau}$	and $R_{\tau\mu}$	\cite{Ledroit}. They are
\begin{equation}\label{fourfermionoperators}
	a^s_{ll}[ab]=\frac{y_b^2}{4m_E^2} \qquad a^s_{ll}[ac]=\frac{y_c^2}{4m_E^2},
\end{equation}
where $y_b$ and $y_c$ are the Yukawa coupling of leptons $b$ and $c$ respectively. Note that being proportional to lepton Yukawa, the contributions to these operators increases as $v_a$ is lowered and will lead to a lower (upper) bound on $v_a$ ($\tan \beta$).

\begin{figure}[tbp]
		\begin{center}
			\includegraphics[scale=0.75, draft=false, trim=0.0cm 0.0cm 0.0cm 0cm, clip=true]{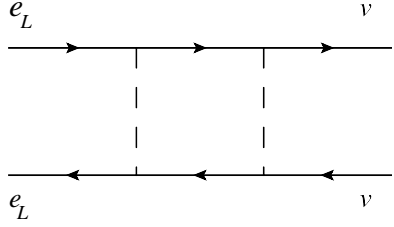}
		\end{center}
		\caption{\label{Boxdiagram} Diagram contributing to the four-fermion operators. Unidentified lines correspond to unspecified superpartners.}
\end{figure}

Box diagrams like those of [\ref{Boxdiagram}] are also accounted for with four-fermion operators, though the expressions are too long to be included here. In the case of diagrams including an exchange of sleptons $a$, the limit of large $\tan{\beta}$ is also taken, as it is strongly overshadowed by mixing effects anyway.

\subsection{Vertex modifying operators}
The operators of \eqref{operators3} receive contributions from lepton mixing, oblique parameters ($W$ and $Y$) and loop diagrams. 
Because the left-handed parts of the leptons of family $a$ mix with superpartners of different gauge charges, the interactions between the gauge bosons and the physical leptons are modified. The result is readily obtained by replacing the leptons of family $a$ and the relevant superpartners by their equivalent combinations of mass eigenstates in the gauge boson vertex interaction terms. This can be summarized in terms of effective operators with coefficients
\begin{equation}
\label{treevertex}
		a^t_{hl}[a]=\frac{1}{8v^2}\left[\left(\frac{gv_a}{M_2^D}\right)^2-\left(\frac{g'v_a}{M_1^D}\right)^2\right],\qquad
		a^s_{hl}[a]=\frac{1}{8v^2}\left[3\left(\frac{gv_a}{M_2^D}\right)^2+\left(\frac{g'v_a}{M_1^D}\right)^2\right].
\end{equation}
The right handed part of the lepton of family $a$ mixes as well, but this effects is suppressed with respect to left-handed mixing by a factor of $(m_a/M_2^D)^2$ and is therefore ignored. These operators lead to deviations of the coupling of the lepton of flavour $a$ to the SM gauge bosons, and will lead to an upper (lower) bound on $v_a$ ($\tan \beta$). 

The contribution from oblique parameters that arise once $O_Y$ and $O_W$ are eliminated are given by
\begin{equation}
	a^s_{hi}=-Y_ig'^2\frac{Y}{4M_W^2} \qquad a^t_{hj}=-g^2\frac{W}{8M_W^2}. 
\end{equation}
Once again, the family indices are suppressed because they are the same for every generation.

\begin{figure}[tbp]
		\begin{center}
			\includegraphics[scale=0.6, draft=false, trim=0.0cm 0.0cm 0.0cm 0cm, clip=true]{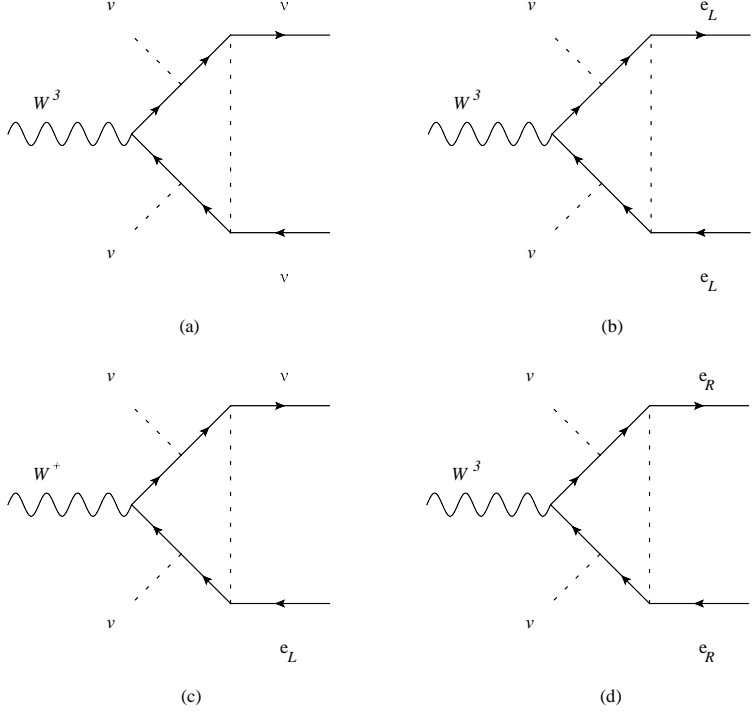}
		\end{center}
		\caption{\label{Triangle diagrams} Corrections to gauge interactions arising from loops. Unidentified lines correspond to unspecified superpartners or leptons $a$. There are four possible types and similar diagrams exist for quarks.}
\end{figure}

Finally, loop-level diagrams of the type shown in figure $[$\ref{Triangle diagrams}$]$ also lead to corrections to the gauge boson fermion vertices. They appear in the four possible ways shown. We can parametrize their effects in terms of \footnote{To avoid a possible sign confusion, we mention that covariant derivatives are taken with a $+$ sign. For example, $D_{\mu}E=(\partial_{\mu}+ig'Y_E B_{\mu})E$.}
\begin{equation}
	\begin{aligned}
		&A_1=-2\sum\mbox{diagrams of type (a)}\\
		&A_2=2\sum\mbox{diagrams of type (b)}\\
		&A_3=-\sqrt{2}\sum\mbox{diagrams of type (c)}\\
		&A_4=2\sum\mbox{diagrams of type (d)}.
	\end{aligned}
\end{equation}
It can be shown that these quantities are related by $A_1+A_2=2A_3$. Using this relation, the diagrams can be accounted for by using the operators of \eqref{operators3}. This leads to coefficients of
\begin{equation}
		a^s_{hl}=\frac{A_2-A_3}{gv^2}\qquad a^t_{hl}=\frac{A_3}{gv^2} \qquad a_{he}=\frac{A_4}{gv^2}.
\end{equation}
The same equations also apply to quarks. The diagrams containing Yukawa coupling are neglected as these are considerably smaller. The limit of large $\tan{\beta}$ is also taken when an internal slepton $a$ is present.

\subsection{Loop contributions to $O_{\hat{W}}$}
The contributions to the operator $O_{\hat{W}}$ arise at loop level. As this operator leads only to triple gauge boson interactions, it is very poorly constrained and its effect on the fit is negligible. It is therefore not included.

\section{Method and results}
\begin{table}[ht!]
	\begin{center}
	\resizebox{\columnwidth}{!}{%
  \begin{tabular}{ |c|c|c|c| }
		\hline
    {} 								& Standard Notation 								& Measurement 																									& Reference				\\ 	
    \hline
    Atomic parity 		& $Q_W(Cs)$ 												&	Weak charge in Cs																							& \cite{Wood:1997zq}				\\
		violation					& $Q_W(Tl)$ 												&	Weak charge in Tl																							& \cite{Vetter:1995vf,Edwards:1995zz}				\\
		\hline
		DIS								& $g_L^2$ , $g_R^2$ 								& $\nu_{\mu}$-nucleon scattering from NuTeV											& \cite{Zeller:2001hh}				\\
		{}								& $R^{\nu}$													& $\nu_{\mu}$-nucleon scattering from CDHS and CHARM						& \cite{Blondel:1989ev,Allaby:1986pd}		\\
		{}								& $\kappa$													& $\nu_{\mu}$-nucleon scattering from CCFR											& \cite{McFarland:1997wx}				\\
		{}								& $g_V^{\nu e}$, $g_A^{\nu e}$			& $\nu$-$e$ scattering from CHARM II														& \cite{Vilain:1994qy}				\\
		\hline
		Z-pole						& $\Gamma_Z$												& Total Z width																									& \cite{LEP:2003aa}				\\
		{}								& $\sigma_h^0$											& $e^+e^-$ hadronic cross section at Z pole											& \cite{LEP:2003aa}				\\
		{}								& $R^0_f(f=e,\mu,\tau,b,c)$					& Ratios of decay rates																					& \cite{LEP:2003aa}				\\
		{}								& $A^{0,f}_{FB}(f=e,\mu,\tau,b,c)$	& Forward-backward asymmetries																	& \cite{LEP:2003aa}				\\
		{}								& $A_f(f=e,\mu,\tau,b,c)$						& Polarized asymmetries																					& \cite{LEP:2003aa}				\\
		\hline
		Fermion pair			& $\sigma_f(f=q,\mu,\tau)$					& Total cross section for $e^+e^- \to f\overline{f}$						& \cite{LEP:2003aa}				\\
		production at			& $A^{f}_{FB}(f=e,\mu,\tau,b,c)$		& Forward-backward asymmetries for $e^+e^- \to f\overline{f}$ 	& \cite{LEP:2003aa}				\\
		LEP2							& $d\sigma_e / d \cos \theta$				& Differential cross section for $e^+e^- \to e^+e^-$  					& \cite{Abbiendi:2003dh}				\\
		\hline
		$W$ pair					& $d\sigma_W / d \cos \theta$				& Differential cross section for $e^+e^- \to W^+W^-$  					& \cite{Achard:2004zw}				\\
		\hline
		{}								& $M_W$															& $W$ mass																					  					& \cite{LEP:2003aa,Abazov:2003sv}		\\
		\hline
		Ratio of	lepton 	& $R_{\tau}$												& Ratio of decay rate of $\tau$ to $e$ on $\tau$ to $\mu$				& \cite{Ledroit}	\\
		decay rate				& $R_{\tau\mu}$											& Ratio of decay rate of $\tau$ to $\mu$ on $\mu$ to $e$				& \cite{Ledroit}	\\
		\hline
	\end{tabular}%
	}
	\caption{Relevant observables. Taken and expanded from \cite{Han:2004az}.}\label{Observables}
	\end{center}
\end{table}
The coefficients of the different operators are constrained using the observables of table [\ref{Observables}]. They are those of \cite{Han:2004az} with a few minor differences. The observable $\sin \theta_{\mbox{\small{eff}}}$ is not used because it assumes lepton universality which is not the case with the model. The ratios of decay rates $R_{\tau}$	and $R_{\tau\mu}$	\cite{Ledroit} are included, as they affect strongly the lower limit on $v_a$. These two observables would be unaffected if $U(3)^5$ symmetry was assumed.

\begin{figure}
        \centering
        \begin{subfigure}[b]{0.425\textwidth}
                \centering
                \includegraphics[width=\textwidth]{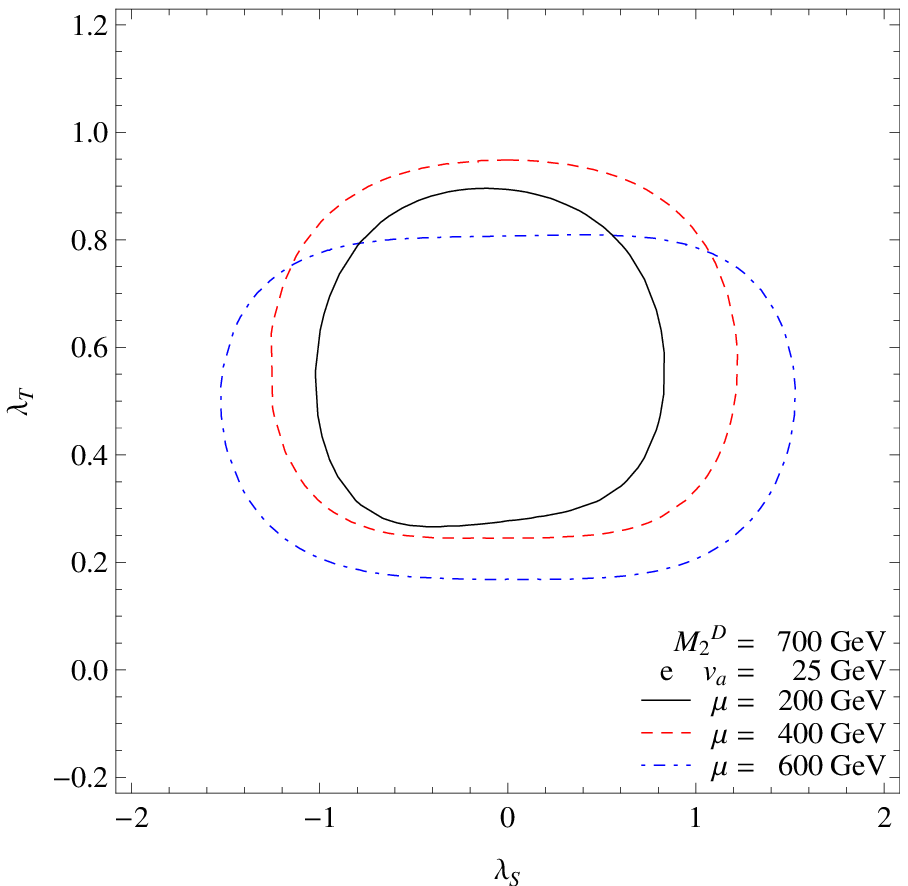}
                \caption{}
                \label{Plot1a}
        \end{subfigure}%
        ~
        \begin{subfigure}[b]{0.425\textwidth}
                \centering
                \includegraphics[width=\textwidth]{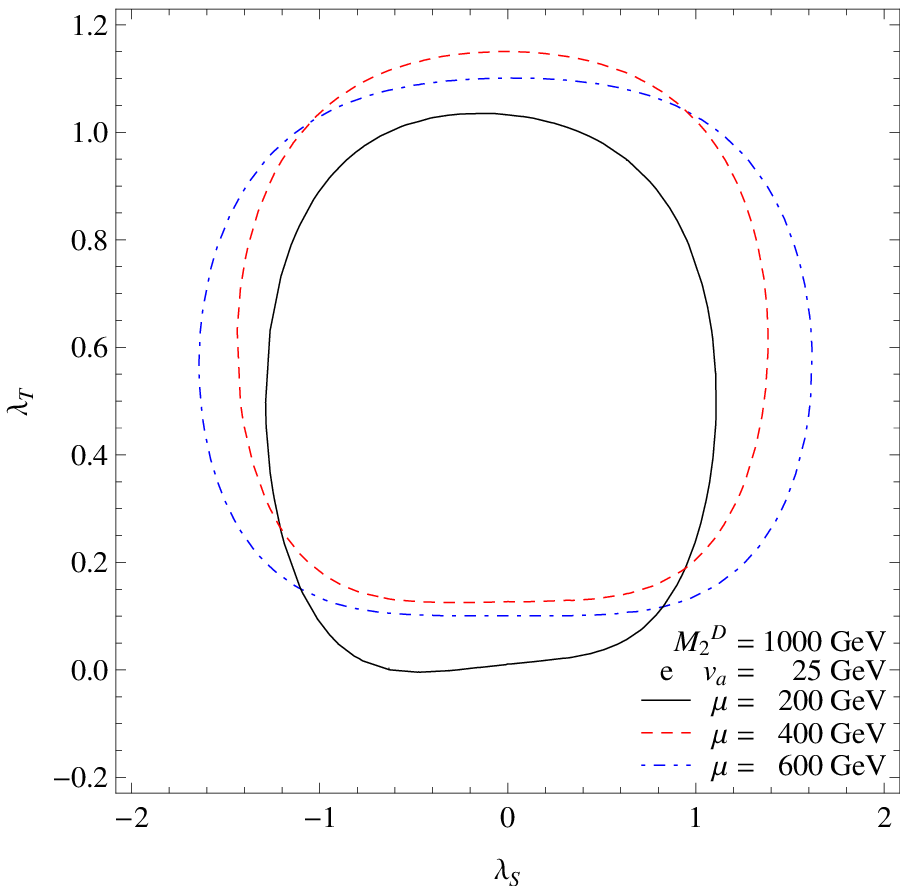}
                \caption{}
                \label{Plot1b}
        \end{subfigure}
				\caption{Region of allowed phase space for the mass combinations of table [\ref{tablemasses}] and for different values of $\mu$. The solid, dashed and dotdashed lines correspond respectively to 200, 400 and 600 GeV. (a) is taken for $M_1^D=M_2^D=700$ GeV and (b) for $M_1^D=M_2^D=1000$ GeV. Both are taken at $v_a=25$ GeV and lepton $a$ is the electron.	The contours correspond to $95.45\%$ confidence level.}\label{Plots1}
\end{figure}

\begin{figure}
        \centering
        \begin{subfigure}[b]{0.425\textwidth}
                \centering
                \includegraphics[width=\textwidth]{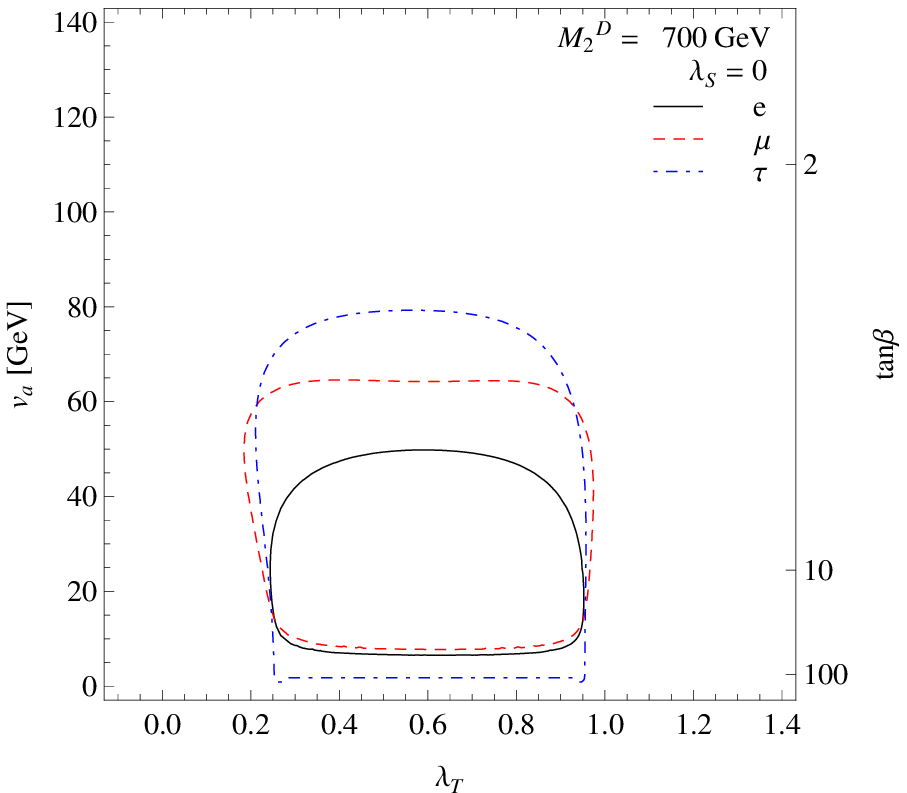}
                \caption{}
                \label{Plot2a}
        \end{subfigure}%
        ~
        \begin{subfigure}[b]{0.425\textwidth}
                \centering
                \includegraphics[width=\textwidth]{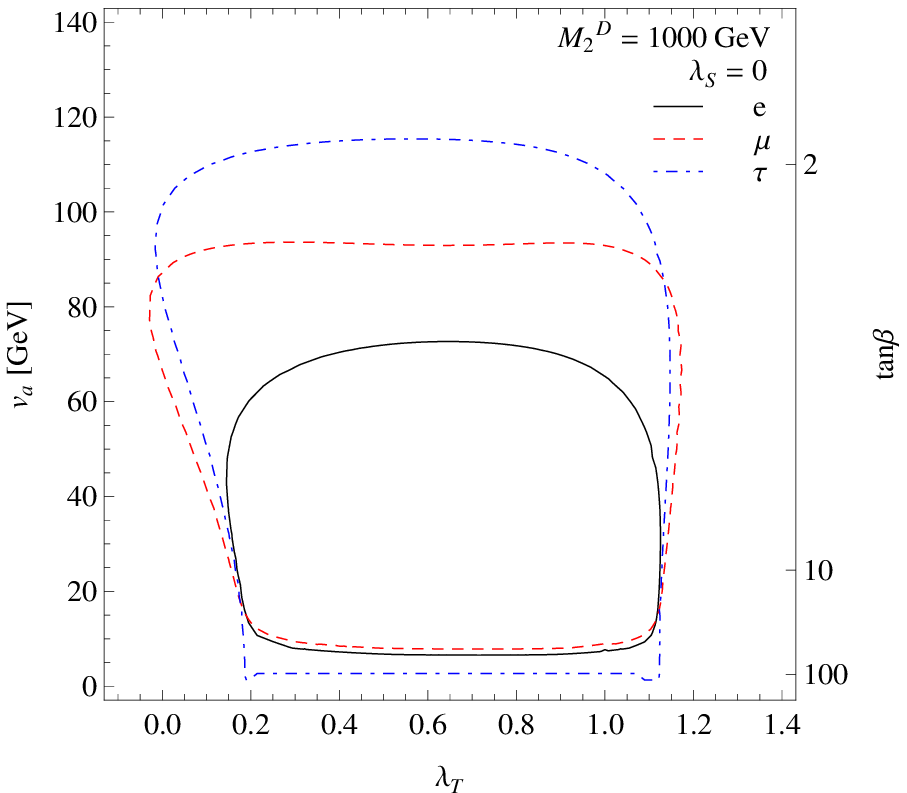}
                \caption{}
                \label{Plot2b}
        \end{subfigure}
				\caption{Region of allowed phase space for the mass combinations of table [\ref{tablemasses}] and for different choices of lepton $(a)$ The solid, dashed and dotdashed lines correspond respectively to the electron, muon and tau. Both figures have $\lambda_S=0$ and (a) has $M_1^D = M_2^D = 700$ GeV  and (b) has $M_1^D=M_2^D= 1000$ GeV. The contours correspond to $95.45\%$ confidence level.}\label{Plots2}
\end{figure}

\begin{figure}
        \centering
        \begin{subfigure}[b]{0.425\textwidth}
                \centering
                \includegraphics[width=\textwidth]{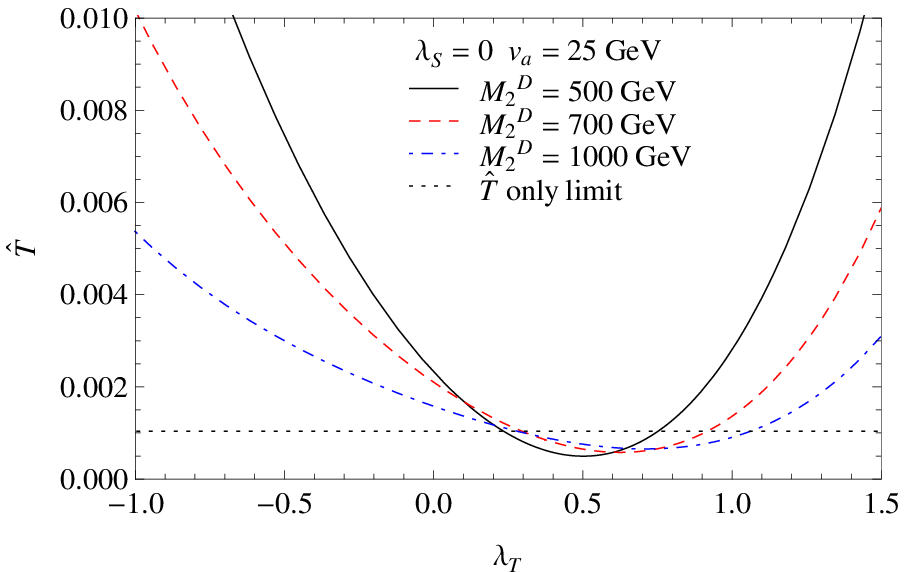}
                \caption{}
                \label{Plot3a}
        \end{subfigure}%
        ~
        \begin{subfigure}[b]{0.445\textwidth}
                \centering
                \includegraphics[width=\textwidth]{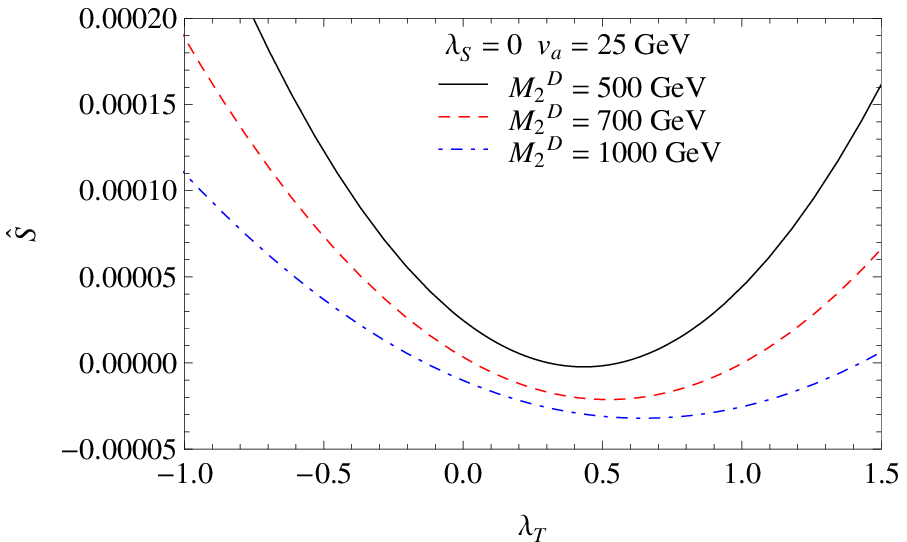}
                \caption{}
                \label{Plot3b}
        \end{subfigure}
				\caption{$\hat{T}$ and $\hat{S}$ parameters as a function of $\lambda_T$ for the masses of table  [\ref{tablemasses}] and different values of $M_2^D$. The solid, dashed and dotdashed lines correspond respectively to 500, 700 and 1000 GeV. Both are taken at $v_a=25$ GeV, $M_1^D=M_2^D$ and with $\lambda_S=0$. The horizontal line corresponds to what the limit on $\hat{T}$ would be at $95.45\%$ if only its corresponding operator would be present and $Y=0$. It's value is $1.04$ $\times$ $10^{-3}$.}\label{Plots3}
\end{figure}

The correction to each observable is calculated to linear order in the coefficients of the higher dimensional operators and this is then used to calculate the $\chi^2$ distribution. Each coefficient is then replaced by its expression in terms of the parameters of the theory. It is then possible to set a number of parameters and do a fit on the remaining ones. These parameters are then constrained inside a region of phase space with a given confidence level.
\begin{table}[ht!]
	\begin{center}
	\resizebox{\columnwidth}{!}{%
  \begin{tabular}{ |c|c|c|c|c|c|c|c|c| }
		\hline
    Figure	& $a$ & $M_{sfermions}$	& $m_{R_d}$	& $\mu$		& $m_{T}$, $m_{S}$ & $M_2^D$, $M_1^D$ & $\lambda_S$ & $\chi^2/\mbox{D.O.F.}$		\\ 	
    \hline
		{}									&	{}			&	{}		& {} 		&	200		& {} 		& {} 				&	{}				&	$0.916$  	\\
    $[$\ref{Plot1a}$]$	&	$e$			&	500		& 700 	&	400		& 1000	& 700				&	Variable	&	$0.883$		\\
		{}									&	{}			&	{}		& {} 		&	600		& {} 		& {} 				&	{}				&	$0.878$  	\\
		\hline
		{}									&	{}			&	{}		& {} 		&	200		& {} 		& {} 				&	{}				&	$0.909$  	\\
    $[$\ref{Plot1b}$]$	&	$e$			&	500		& 700 	&	400		& 1000	& 1000			&	Variable	&	$0.883$		\\
		{}									&	{}			&	{}		& {} 		&	600		& {} 		& {} 				&	{}				&	$0.879$  	\\
		\hline
		{}									&	$e$			&	{}		& {} 		&	{}		& {} 		& {} 				&	{}				&	$0.879$  	\\
    $[$\ref{Plot2a}$]$	&	$\mu$		&	500		& 700 	&	400		& 1000	& 700				&	0					&	$0.881$		\\
		{}									&	$\tau$	&	{}		& {} 		&	{}		& {} 		& {} 				&	{}				&	$0.880$  	\\
		\hline
		{}									&	$e$			&	{}		& {} 		&	{}		& {} 		& {} 				&	{}				&	$0.881$  	\\
    $[$\ref{Plot2b}$]$	&	$\mu$		&	500		& 700 	&	400		& 1000	& 1000			&	0					&	$0.882$		\\
		{}									&	$\tau$	&	{}		& {} 		&	{}		& {} 		& {} 				&	{}				&	$0.881$  	\\
		\hline
		{}									&	{}			&	{}		& {} 		&	{}		& {} 		& 500				&	{}				&	{}		  	\\
    $[$\ref{Plot3a}$]$, $[$\ref{Plot3b}$]$	&	N$\slash$A			&	500		& 700 	&	400		& 1000	& 700				&	0		&	N$\slash$A			\\
		{}									&	{}			&	{}		& {} 		&	{}		& {} 		& 1000			&	{}				&	{}		  	\\
		\hline
	\end{tabular}%
	}
	\caption{Masses for each plot in GeV and choice of lepton $a$. $\chi^2/$ D.O.F. stands for the best fit of $\chi^2$ divided by the number of degrees of freedom. For the standard model, $\chi^2/$ D.O.F is $0.883$. In all cases, $b_T=b_S=0$.}\label{tablemasses}
	\end{center}
\end{table}
For the set of masses that we consider (shown in table [\ref{tablemasses}]), we find that fits of our model to the data are roughly as good as the Standard Model fit. As mentioned above, the {\it vev} for the triplet $T_3$ is potentially dangerous so we restricted the mass parameter of the scalar triplet to be 1 TeV. The result of the $\chi^2$ per degree of freedom for different choices of parameters is also shown in table [\ref{tablemasses}]. The allowed region of parameter space, in the $\lambda_S$/$\lambda_T$/$v_a$ space consists of a roughly cylindrical region whose flat sides are parallel to the $\lambda_T /\lambda_S$ plane. The $\hat{T}$ parameter has contributions that scale as $\lambda_T^4$ and $\lambda_S^4$ and is the dominant factor in determining the shape of the allowed region for  $\lambda_T$ and  $\lambda_S$. Figure [\ref{Plots1}] shows this region of allowed phase space in the $\lambda_S\slash \lambda_T$ plane for different combinations of masses which can be found in table [\ref{tablemasses}].  We show the 95.45$\%$ confidence level exclusion only as the lines of confidence level of 68.27$\%$ and 99.76$\%$ are very close to those of 95.45$\%$ and  provide little new information. We see that there are strong bounds on those parameters, with the allowed region becoming larger as  $M_{(1,2)}^D$ is increased as it set the scale of $v_{T^3}$ and the masses of the fermions which give the largest loop contributions to $\hat{T}$. Overall, we see  that $\lambda_S$ and $\lambda_T$  cannot take values much greater than one irrespective of the masses of the superpartner or the choice of generation for the lepton $a$. This has important consequences for radiative corrections to the Higgs mass in this model \cite{Bertuzzo:2014}.  
 As explained before, contributions \eqref{fourfermionoperators} and \eqref{treevertex} lead to a two-sided bound on $v_a$ which is constrained to be rather small, though the exact range depends considerably on the choice of masses and the generation of the lepton $a$. This is illustrated in figure [\ref{Plots2}] which shows the allowed region in the $ \lambda_T \slash v_a$ plane. Again, the upper bound on $v_a$ is relaxed as $M_{(1,2)}^D$ is increased and can reach a value where $\tan \beta = 2$ for $a=\tau$ and $M_{(1,2)}^D = 1000$ GeV. The lower bound on $v_a$ primarily depends on the mass of the sfermions which are not varied in the figures. 
In  figure [\ref{Plots3}], we show the  $\hat{T}$ and $\hat{S}$ parameters as a function of $\lambda_T$. We see that the $\hat{S}$ parameter is very small while the $\hat{T}$ parameter can become sizable and drives the limit on $\lambda_T$.  

Overall, as one would expect, increasing the various mass parameters will relax the various bounds. The situation for $\mu$ is however slightly more involved as increasing $\mu$ will increase $v_{T_3}$ (see equation \eqref{vevT3}) which can then be taken back to an acceptable value by constraining $\lambda_T$ to be close to $\lambda_T  \sim g M_2^D(v_u^2-v_a^2)/(\sqrt{2} \mu v_u^2)$.

\section{Conclusions}
In this paper, we have studied the bounds that electroweak precision measurements put on a supersymmetric model where a lepton number is identified with a  $U(1)_R$ symmetry and one of the sneutrino gets a {\it vev} and is responsible for giving  masses to the down-type quarks and the leptons. Deviations from Standard Model predictions for electroweak precision observables come from various sources: mixing of one of the lepton flavor with the gauginos, a {\it vev} for an $SU(2)$ scalar triplet, tree level exchange of sfermions through the Yukawa couplings and loops of superpartner. To bound the parameter space of this model we used the higher dimensional operators method developed in \cite{Han:2004az}, extended to include the relevant breaking of lepton universality. Our results are illustrated in figure [\ref{Plots1}] and [\ref{Plots2}]. The first figure shows that $\lambda_T$, the superpotential coupling between the Higgs doublets and the $SU(2)$ triplet, is prevented from taking large value. This coupling also gives a sizable radiative contribution to the Higgs mass so this bound has important repercussions \cite{Bertuzzo:2014}. However, note that in this work we remained agnostic about the precise mechanism that gives the Higgs its correct mass. The second figure shows the bounds on the {\it vev} of the sneutrino. It cannot be too large as this creates a large mixing between the corresponding lepton and the gaugino, but it also cannot be too small as this implies a large $\tau$ Yukawa coupling, and leads to deviation to $R_{\tau}$ and $R_{\tau/\mu}$. Nevertheless, a relatively large range is still allowed, with the possibility of having $\tan \beta$ as low as $2$ for gauginos with a 1 TeV Dirac mass. 

\acknowledgments
We would like to thank Enrico Bertuzzo, Claudia Frugiuele and Eduardo Pont\'on for useful discussions. This work was supported in part by the Natural Sciences and Engineering Research Council of Canada (NSERC).


\bibliographystyle{JHEP}
\bibliography{Draft9}

\end{document}